# Intelligent Bandwidth Allocation for Latency Management in NG-EPON using Reinforcement Learning Method

Qi Zhou[1, *], Jingjie Zhu[2], Junwen Zhang[2, *], Zhensheng Jia[2], Bernardo Huberman[2,*], and Gee-Kung Chang[1]

[1]Georgia Institute of Technology, Atlanta, GA; [2]CableLabs, Inc., Louisville, CO 80021 USA

Email address: qi.zhou@gatech.edu; j.zhang@cablelabs.com; b.huberman@cablelabs.com

**Abstract:** A novel intelligent bandwidth allocation scheme in NG-EPON using reinforcement learning is proposed and demonstrated for latency management. We verify the capability of the proposed scheme under both fixed and dynamic traffic loads scenarios to achieve <1ms average latency. The RL agent demonstrates an efficient intelligent mechanism to manage the latency, which provides a promising IBA solution for the next-generation access network.

## 1. Introduction

Delivering more bandwidth/capacity has been a top research focus in optical access networks. Moreover, new services such as 5G mobile X-haul, edge computing, AR/VR Gaming, Tactile Internet and UHD video distribution, are placing additional requirements on access networks [1]. This implies the importance of low latency and high reliability will be increasing for future access networks, as they will be asked to deliver time-critical services. Thus, new deterministic and reliable latency management approaches are necessary. For instance, a 1-10 ms e2e latency is required for the F1 mobile fronthaul interface, while this number reduces to only several hundreds of μs (<1ms) if we move to a lower layer function-split of mobile fronthaul [1].

As a point-to-multi-point system, Passive Optical Network (PON) has been one of the dominant architectures to provide bandwidth sharing among different types of services [2]. In general, dynamic bandwidth allocation (DBA) is used in PON to allocate traffic bandwidth in upstream based on the instantaneous demands and requests from users (ONUs). Different DBA algorithms or strategies have been proposed to support the upstream bandwidth sharing [3], however, most of the algorithms are based on a fixed strategy and have no feedback from the network environment changes and use-scenario requirements upgrading. Theoretically, different DBA algorithms would be suitable for different use-scenarios or traffic conditions. In addition, the "optimal" DBA scheme for the same network can vary from time to time as the traffic load during a day or week can change dramatically. In addition, different users/services may have distinct latency requirements. When the traffic load from each user changes, the corresponding network latency also changes. As mentioned above, many emerging services require more deterministic and reliable latency. Therefore, an intelligent bandwidth allocation that can perceive or sense the network environment changes proactively and correspondingly updates its bandwidth allocation policy smartly to manage the latency for different users would be very attractive. Recently, machine-learning-based methods have been reported in bandwidth and resource allocation in wireless and mobile access networks, which show promising performances [4].

In this paper, we propose a novel method for intelligent bandwidth allocation (IBA) in PON by using reinforcement learning (RL) for latency management. The reinforcement learning scheme consists of three core factors, including *State, Action,* and *Reward*, which are implemented to proactively update the bandwidth allocation parameters. Based on the network traffic information as the input State, the core Agent updates the maximum allocated bandwidths of the target user for latency management. We verify the capability of the proposed scheme under both fixed and dynamic traffic loads scenarios to achieve <1ms average latency. The RL agent demonstrates an efficient intelligent mechanism to manage the latency, which provides a promising IBA solution for the next-generation access network.

## 2. Principle of Reinforcement Learning for IBA in PON

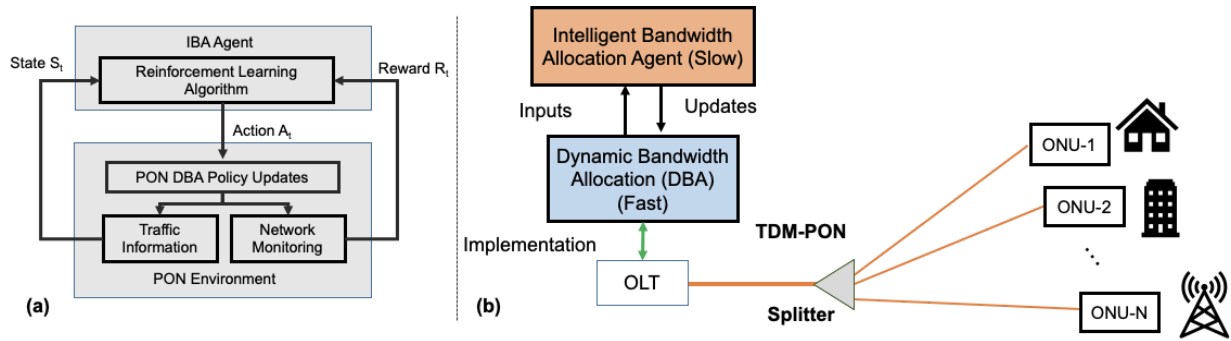

Fig. 1. The Principle and implementation of RL for IBA in PON: (a) the principle; (b) the implementation in the PON environment.

Figure 1 shows the principle and implementation of reinforcement learning (RL) for IBA in PON, and Fig. 1(a) shows the detailed processing flow, and Fig. 1 (b) shows the implementation in the PON environment. RL has demonstrated prominent performance in strategy selection and optimization tasks, such as AlphaGo [5], professional gaming [6] and interference avoidance [7], to name a few. The RL agent can obtain positive/negative rewards on its executed action under a certain state through interaction with the environment. The feedbacks on state-action pairs can be saved and updated using Q-table or deep neural network, such that the agent is able to make the decision with the most positive expected reward. For our demonstration, the input State $S_t$ is the traffic information as the average load of traffic in ONU over the time, while the Action is the optimal maximum bandwidth $W_{max}(i)$ allocated to the $i_{th}$ ONU, which is a series of discrete values $W(n) \in [W_1, …, W_N]$. The Q-table updating and action selection are based on State–action–reward–state–action (SARSA) algorithm, which is an on-policy temporal difference value-based RL algorithm with more conservative action to ensure reliable operations in NG-EPON. Here the *Reward* is the traffic latency of specific ONU to the target latency after an action.

A two-layer implementation of proposed IBA in PON with different time scales is shown in Fig. 1(b). In the lower layer, fast DBA is implemented in μsec to msec scale for real-time bandwidth assignments according to the actual bandwidth request from ONUs. On the higher layer, we have the IBA agent to update the DBA policy in the scale of hundreds of msec to seconds range based on the inputs from the network. The IBA actions are implemented in a non-real-time manner, which is based on the RL algorithm described above.

## 3. Physical Layer Parameters Test

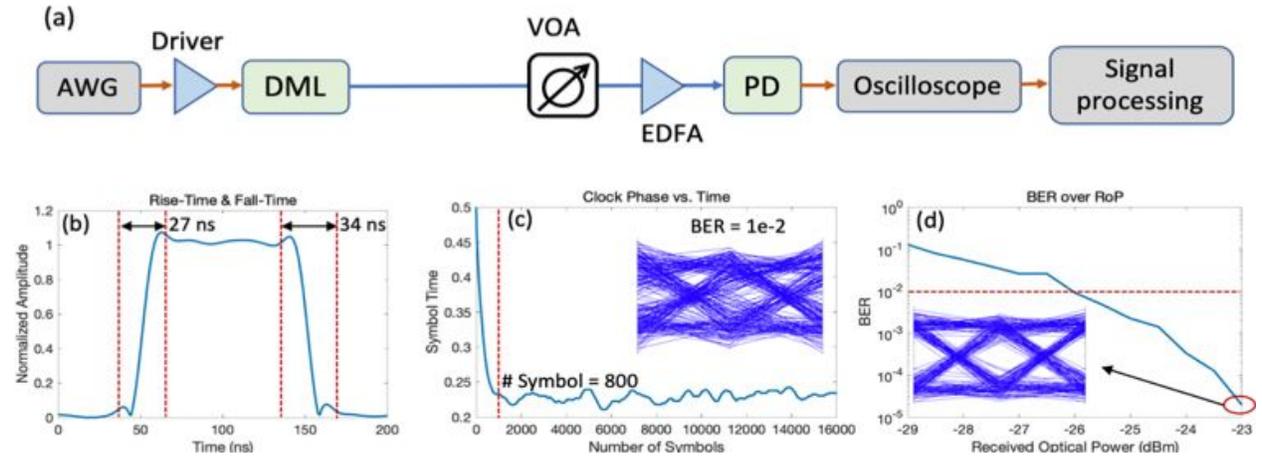

Fig.2(a) Experimental setup; (b) laser on-off time measurement; (c) BCDR performance at BER = 1e-2; (d) BER over received optical power (RoP) measurement.

To find out the key parameter settings of IBA simulation, we conduct physical layer verifications including laser on-off time, burst-clock-data-recovery (BCDR) time and link sensitivity test for 25G EPON. As such, the total effective guard interval time can be confirmed. The experimental setup is shown in Fig.2(a). We first randomly generate a 100k length bit sequence and perform pre-equalization based on estimated channel coefficients derived by the least-mean-square (LMS) algorithm. The pre-equalized sequence is then downloaded to an arbitrary waveform generator (AWG) to produce an NRZ-OOK signal running at 25 Gb/s to comply with the NG-EPON standard. The signal is amplified by a modulator driver and is modulated onto a directly modulated laser (DML) for fiber transmission. The optical signal is detected by a PIN receiver with optical preamplification and is digitized by an oscilloscope for the following off-line signal decoding. A variable optical attenuator (VOA) is implemented to control the received optical power (RoP). Fig.2(b) shows the measured laser on-off behavior, the laser rise time is 27 ns while the laser fall time is 34 ns. Besides, to achieve a line rate of 25 Gb/s in NG-EPON, a fast synchronization and signal recovery using BCDR for the burst-mode upstream is necessary. We implement a first-order digital CDR scheme based on 2-times oversampling and nonlinear bang-bang phase detector [8] to test the BCDR speed. Follow the BER threshold in IEEE 802.3ca specification, Fig.2(c) demonstrates the clock phase vs time result at BER = $1e^{-2}$. The clock phase is converged at around 0.23 symbol time after 800 symbols of training. At a sampling rate of 50 GSa/s, the BCDR convergence time is derived to be 16 ns. Fig.2(d) is the BER over RoP measurement to calculate the link budget of the NG-EPON. The receiver sensitivity is -26 dBm at BER = $1e^{-2}$. By taking the difference between the DML output power of 5.5 dBm and receiver sensitivity, the link budget of the system reaches 31.5 dB. Since there is no commercially available burst-mode transimpedance amplifier (BM-TIA) for burst gain setting time measurement, we calculate the total effective guard interval time based on the gain setting time of 48 ns from the reference work of 25-GBaud BM-TIA [9]. We confirm that considering the laser on-off time, BCDR time as well as the BM-TIA setting time, a total effective guard interval time of 1 μs should be safe enough for future commercial systems. The effective guard time setting of 1 μs will be used in our following simulations.

## 4. Intelligent Bandwidth Allocation Simulations and Results

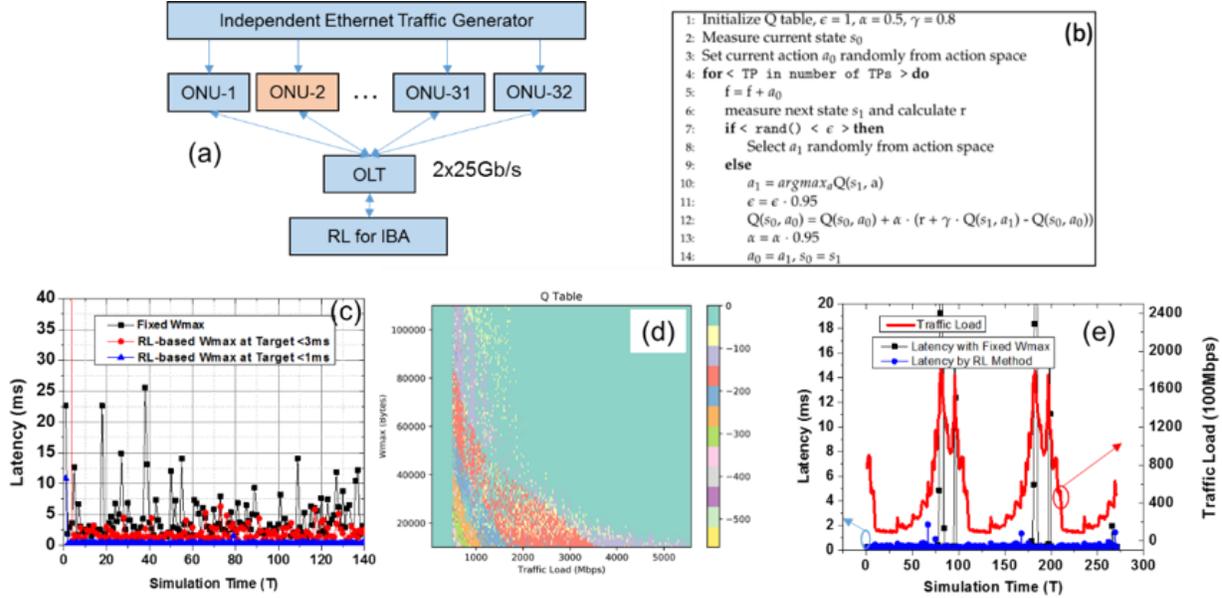

Fig. 3. The simulation setup and results: (a) the simulation setup; (b) the SARSA algorithm used in the simulation; (c) simulation of RL to target 3-ms latency; (d) the Q-table value obtained; (d) the latency performance with dynamic traffic load.

Figure 3 shows the simulation setup and results obtained. We simulated 32 ONUs in the NG-EPON system with two wavelengths each carrying 25 Gb/s data, as shown in Fig. 3 (a). We assume 1 μs for guard interval time according to our experimental verification above. As such, a total of 50-Gb/s capacity is shared by the 32 ONUs using the first-fit scheme for DBA on the two wavelengths. All 32 ONUs have random RTTs within the range of 100 to 200 μs. ONU2 is the target ONU that is enabled with latency management based on the RL method. All traffic is generated by an Ethernet traffic generator model that is described in [10], where self-similar traffic is generated based on the aggregation of multiple streams, each consisting an alternating Pareto-distributed ON/OFF period [10]. The Ethernet traffics is with the packet size of 64 to 1518 bytes, and the maximum traffic load for each ONU is 2 Gb/s. The default $W_{max}$ for simulation is set at 30000 bytes. The Q-table update interval and $W_{max}$ adjustment interval are all set as 0.8 s. The algorithm for the implemented SARSA learning is designed as shown in Fig. 3 (b). Fig. 3 (c) shows the latency management results at the fixed load rate of 1.0 at 2 Gb/s. To verify the latency management capability, we set two target latency values at 3 ms and 1 ms. It is seen that in the result of Fig. 3 (c) the target ONU2 follows the latency targets < 3 ms and < 1 ms with our latency management. As a comparison, we plot the latency performance under a fixed $W_{max}$ setting to present that the variance of latency is significantly reduced by employing latency management. Fig. 3 (d) shows the Q value distribution of the Q-table after training with 1-ms target latency and different traffic loads, we can see that the peak data rate of upstream burst traffic can be as high as 5.5 Gb/s. Finally, the latency management performance with dynamic traffic loads is shown in Fig. 3 (e), with the simulation the traffic load changes based on the trend obtained from a real user traffic behavior during a day. As a comparison, the latency performance of fixed $W_{max}$ is also presented. The latency of fixed $W_{max}$ (30000 bytes) can increase beyond 20-ms at a high traffic load. By using RL method, with a target latency of 1 ms, the determinism and reliability of latency management are demonstrated by the simulation result that the average latency of ONU2 is below 1 ms, with a peak latency about 2 ms due to bursty traffic.

## 5. Conclusions

In this paper, we propose a novel method for intelligent bandwidth allocation in PON by using SARSA reinforcement learning for latency management. The proposed scheme's capability to achieve < 1 ms average latency under both fixed and dynamic traffic loads scenarios is verified. An efficient intelligent mechanism to manage latency is demonstrated by the agent, offering a future-proof IBA solution for NG-EPON.